\documentclass[reprint,
 superscriptaddress,
 amsmath,
 amssymb,
 aps,
 pra,
]{revtex4-1}
\usepackage[utf8]{inputenc}
\usepackage{graphicx}
\usepackage{dcolumn}
\usepackage{bm}
\usepackage{natbib}
\setcitestyle{super}
\usepackage{dirtytalk}
\usepackage{csquotes}
\usepackage{subcaption}
\usepackage{multirow}
\usepackage{textgreek} 
\usepackage{xcolor}

\begin{document}

\title{Dynamic nanoscale spatial heterogeneity in a perovskite to brownmillerite topotactic phase transformation}

\author{Nicolò D'Anna}
\email{ndanna@ucsd.edu}
\affiliation{Department of Physics, University of California San Diego, La Jolla, CA 92093, USA}
\author{Erik S. Lamb}
\affiliation{Department of Physics, University of California San Diego, La Jolla, CA 92093, USA}
\author{Robin Glefke}
\affiliation{Department of Physics, University of California San Diego, La Jolla, CA 92093, USA}
\author{Daseul Ham}
\affiliation{Pohang Light Source-II (PLS-II) Beamline Department, Pohang Accelerator Laboratory, POSTECH, Pohang 37673, Republic of Korea}
\author{Ishmam Nihal}
\affiliation{Advanced Light Source, Lawrence Berkeley National Laboratory, Berkeley, California 94720, USA}
\author{Su Yong Lee}
\affiliation{Pohang Light Source-II (PLS-II) Beamline Department, Pohang Accelerator Laboratory, POSTECH, Pohang 37673, Republic of Korea}
\author{Yayoi Takamura}
\affiliation{Department of Materials Science and Engineering,University of California Davis, Davis, California 95616, USA}
\author{Oleg G. Shpyrko}
\affiliation{Department of Physics, University of California San Diego, La Jolla, CA 92093, USA}

\begin{abstract}
\begin{center}
     \noindent\textbf{Abstract} 
\end{center}
\textbf{Phase transitions are omnipresent in modern condensed matter physics and its applications. 
In solids, first-order phase transformations typically occur by nucleation and growth under non-equilibrium conditions. Under constant external conditions, \textit{e.g.}, constant annealing temperature and pressure, the nucleation and growth dynamics are often thought of as spatially and temporally independent.
Here, \textit{in-situ} Bragg \mbox{X-ray} photon correlation spectroscopy (XPCS) reveals nanoscale spatial and dynamical heterogeneity in the perovskite-to-brownmillerite topotactic phase transformation in La$_{0.7}$Sr$_{0.3}$CoO$_3$ thin films annealed under constant reducing conditions over a time span of multiple hours.
Specifically, a timescale associated with domain growth remains stable, with a corresponding domain wall speed of $v_d = 6 \pm 0.5 \times10^{-4}$~nm/s ($2 \pm 0.2$~nm/h), while a slower timescale, associated with temperature-driven de-pinning of domains, leads to accelerating dynamics with timescales following an aging power law with exponent -2.2$\pm$0.5.
This experiment demonstrates that Bragg XPCS is a powerful tool to study nanoscale dynamics in structural phase transformations, with the ability to extract quantitative average values related to nano-domain motion \textit{in-situ}. The results are relevant for phase engineering of phase-change devices, as they show that nanoscale dynamics, linked to domain and domain-wall motion, can continuously evolve and speed up with time, even hours after the initiation of the phase transformation, with potential repercussions on electrical performance.}
\end{abstract}

\maketitle


\begin{figure*} 
\centering   
    \includegraphics[width=\linewidth]{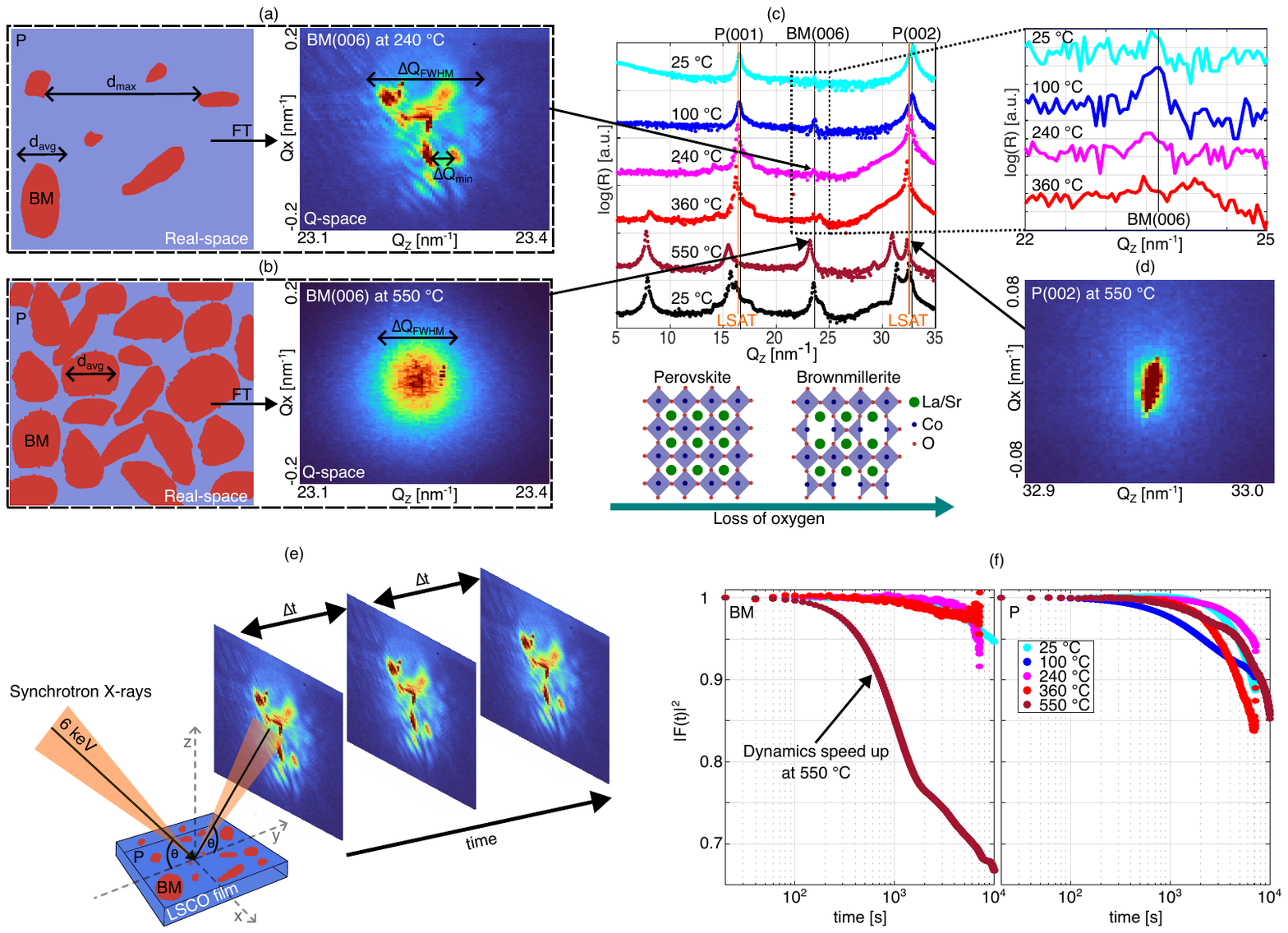}
\caption{\textbf{Coherent x-ray diffraction setup.} 
    BM(006) half-order Bragg diffraction peak (right) observed at 240~$^\circ$C (a) and 550~$^\circ$C (b). Illustrations (left) of a dilute (a) and dense (b) BM phase distribution that can lead to such diffraction patterns. The distance $\Delta Q_{FWHM}$ in Q-space is inversely proportional to the mean in-plane domain \mbox{size $d_{avg}$;} similarly $Q_{min}$ is inversely proportional to the largest distance $d_{max}$ between domains within the X-ray beam.
    (c) XRD measurements throughout the experiment, starting at 25~$^\circ$C (light blue), heating incrementally to 550~$^\circ$C (dark red), and cooling to 25~$^\circ$C (black). The crystal structure of the two observed phases is illustrated on the bottom.
    On the right is a magnification of the BM(006) peak at temperatures below the phase transformation, showing trace amounts of BM phase.
    (d) P(002) Bragg diffraction peak measured at 550~$^\circ$C.
    (e) Illustration of the XPCS experiment.
    (f) Correlation decay of the speckle images measured at all temperatures for the BM(006) peak (left) and the P(002) peak (right), showing a clear increase in dynamics associated with the BM(006) peak at 550~$^\circ$C.
    }
\label{fig_setup}
\end{figure*} 

\noindent\textbf{1. Introduction}\\ 
Aging dynamics, commonly observed in glassy materials, is characterized by dynamical relaxation rates that vary even under constant external \mbox{conditions \cite{angell1995formation,debenedetti2001supercooled,berthier2023modern,d2001jamming}.}
In solids, phase transitions driven by an external stimulus such as temperature, often result in the nucleation and growth of the new phase, forming structural domains that grow until a phase-pure state is \mbox{achieved \cite{li2022situ}.}
In the case of first-order phase transformations where the phases have distinct crystal structures, as is the case for oxygen ion diffusion-driven topotactic phase transformations of metal oxides \cite{gilbert2018ionic,chiu2021cation,d2026jamming}, the evolution of both crystalline phases can be unambiguously detected by Bragg diffraction.
Bragg X-ray photon correlation \mbox{spectroscopy (XPCS)} is a technique particularly well suited to measure slow ($>$1~s) dynamics linked to nanoscale motion of structural domains and domain walls in such diffusion-driven phase transformations. It has the potential to reveal whether nanoscale aging dynamics occurs under constant external conditions on timescales of several hours.
In current efforts to control and utilize phase transitions in phase-change materials for the development of novel technologies such as neuromorphic computing \cite{hoffmann2022quantum} and solid oxide fuel cells \cite{kim2012probing}, understanding the spatial and temporal phase transformation heterogeneity becomes highly relevant.

In the scope of novel energy-efficient technologies beyond semiconductors, a broad interest exists in taking advantage of tunable phase transformations in functional metal oxides \cite{hoffmann2022quantum}. 
Of particular interest are materials where control over the metal-to-insulator transition (MIT) and concomitant structural transition is possible via electric or electrochemical gating \cite{sawa2008resistive,li2018review}. Candidate materials include the cobaltite perovskites \cite{zhang2020understanding,lengsdorf2004pressure},
vanadium oxides \cite{mcwhan1970metal,zylbersztejn1975metal,mcwhan1973metal,adda2022direct,rischau2024resistive}, manganite perovskites \cite{salev2024local,salev2021transverse,chen2024voltage,chen2024electrical}, niobium dioxide \cite{pickett2013scalable,gao2017nbox}, and \mbox{samarium nickelates \cite{shi2013correlated,ha2013electrostatic,rischau2025synaptic}}.
Among such phase-change materials, cobaltite perovskites give access to a particularly rich phase diagram, through a variety of tuning knobs, including epitaxial strain \cite{yin2022compressive,shin2022tunable} and modification of oxygen ion concentration and distribution \cite{caciuffo1999structural,wu2003glassy,chiu2021cation,gilbert2016controllable,jeen2013topotactic}. As a result, their magnetic and electronic properties can be readily manipulated \cite{feng2024hydrogen,chaturvedi2021doping}, thus enabling many applications such as energy conversion \cite{lu2022enhanced}, convolutional neural networks, spintronics, and neuromorphic computing \cite{hoffmann2022quantum,chen2021observation}.
Currently, a focus exists on utilizing topotactic phase transformations in perovskite (P) oxides (chemical formula ABO$_3$), due to their high oxygen conductivity and low oxygen vacancy formation energy which enable phase manipulation through the control of oxygen \mbox{deficiency \cite{jeen2013reversible,jeen2013topotactic,walter2016electrostatic}.}
A number of \mbox{P-related} phases have been identified, such as the Grenier (ABO$_{2.7}$), brownmillerite (BM, ABO$_{2.5}$), square planar (ABO$_2$), and Ruddlesden-Popper (A$_{n+1}$B$_n$O$_{3n+1}$) phases \cite{anderson1993structural}. 
Examples of where the reversible modulation of the oxygen vacancy concentration has been demonstrated in metal oxides include YBa$_2$Cu$_3$O$_{7-\delta}$ \cite{murray2019interfacial},
SrCoO$_3$ \cite{jeen2013topotactic,shih2025freestanding},
SrFeO$_3$ \cite{wang2019brownmillerite,shin1978order},
La$_{0.7}$Sr$_{0.3}$CoO$_3$ \cite{gilbert2018ionic}, and  La$_{0.67}$Sr$_{0.33}$MnO$_3$ \cite{grutter2016reversible}.
\begin{figure*}[ht] 
 \centering   
  \begin{subfigure}[b]{0.48\textwidth}  
  \includegraphics[width=\textwidth]{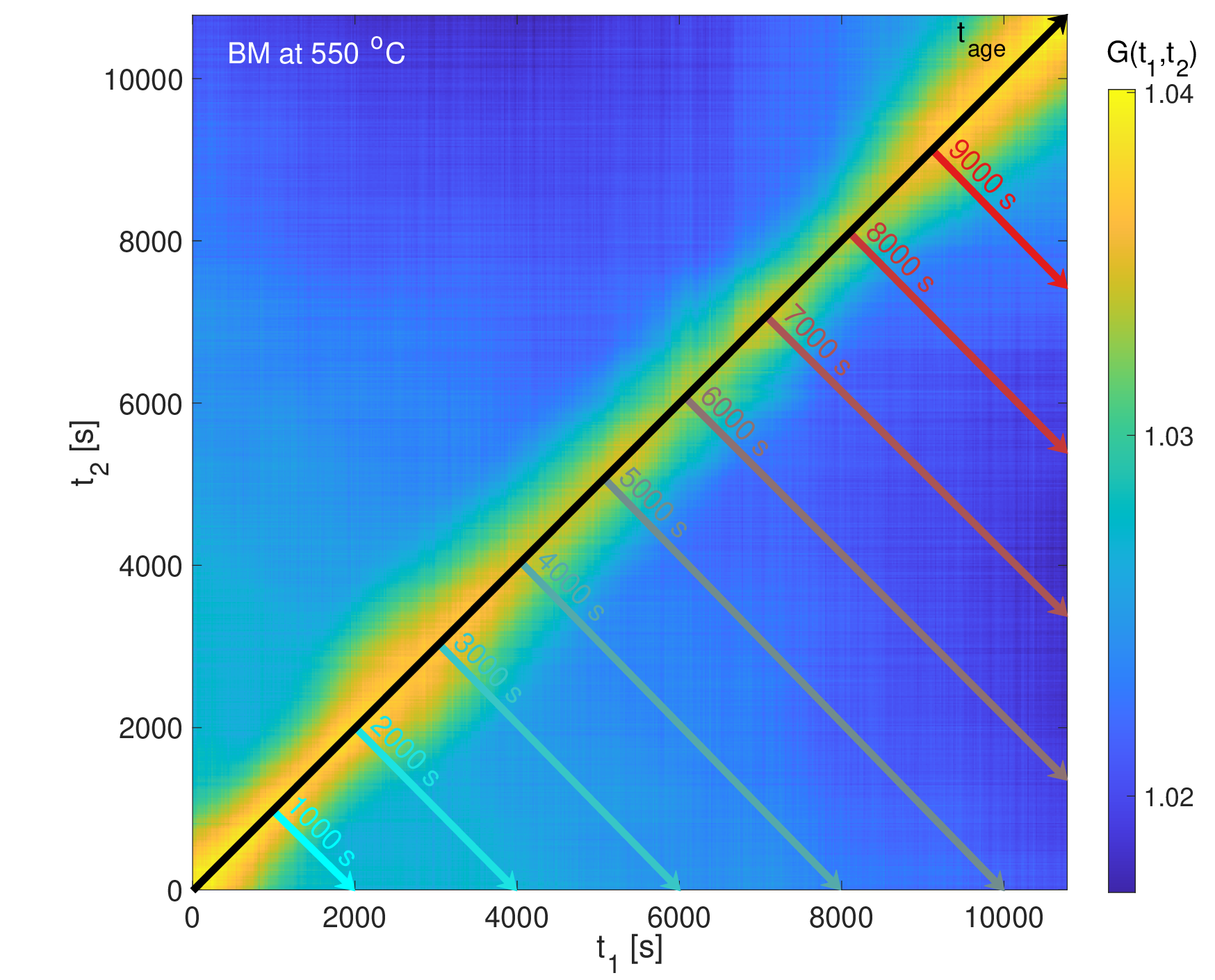}
  \caption{\label{fig_twotimes}}
  \end{subfigure}
\hfill
  \begin{subfigure}[b]{0.48\textwidth}  
  \includegraphics[width=\textwidth]{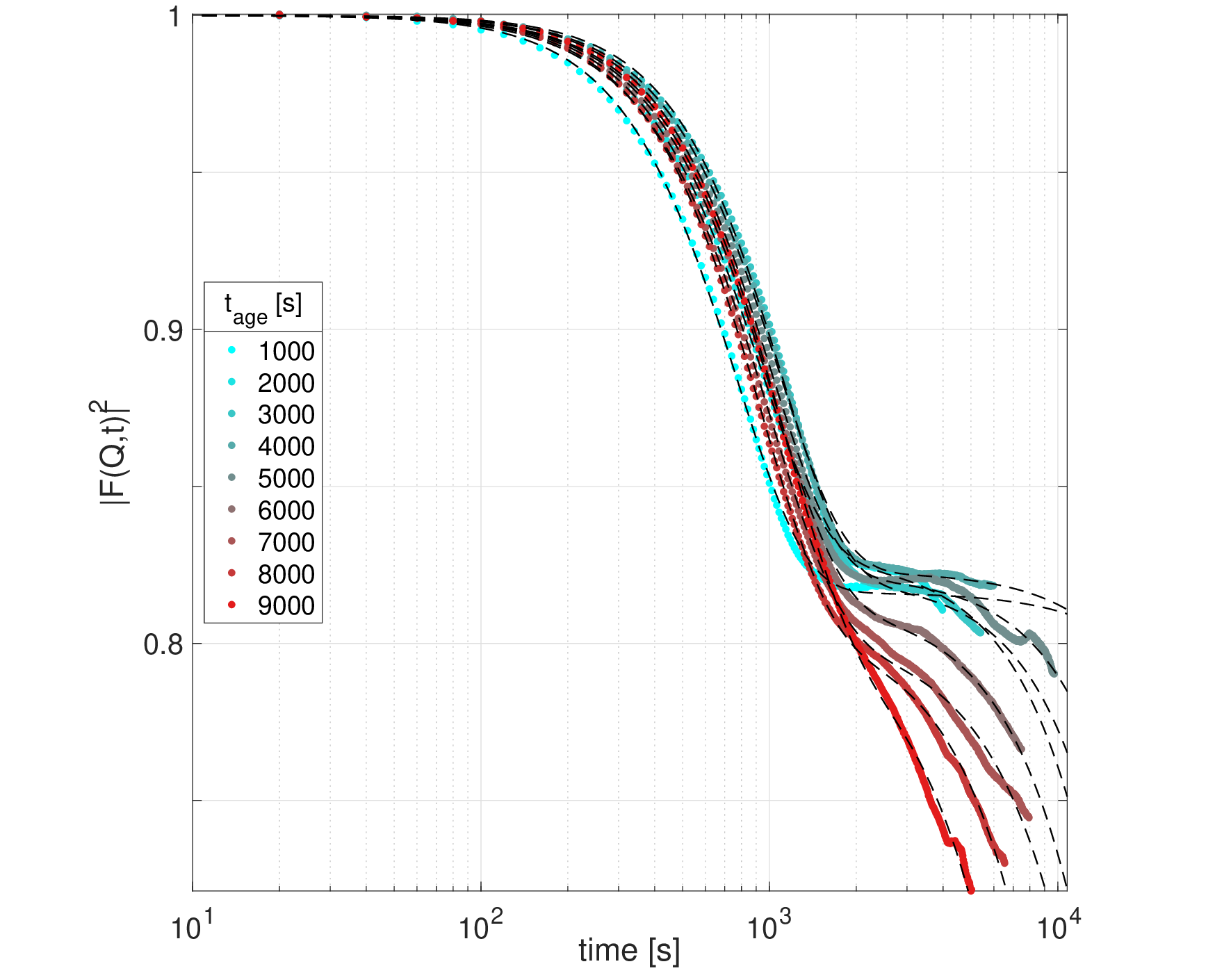}
  \caption{\label{fig_g2_fits}}
  \end{subfigure}
 \caption{\textbf{Dynamic heterogeneity of the phase transformation.} 
 	(a) Two-time correlation $G($t$_1$,t$_2)$ measured on the BM peak at 550~$^\circ$C. The black arrow indicates the direction of increasing t$_{age}$ and follows $G($t$_1$,t$_2$=t$_1)$. Perpendicular colored arrows indicate the data used to calculate the one-time correlation functions $g_2^{t_{age}}$(t) shown \mbox{in (b).}
    (b) t$_{age}$-dependent correlation decay $|F_{t_{age}}$(t)$|^2$ obtained from the $g_2^{t_{age}}$(t) arrows in (a). Dashed lines are fits with the double exponential decay given in Eq.~\ref{exp_fit_eq}.
    }
\label{figtagedef}            
\end{figure*}

Among metal oxides displaying topotactic transitions linked with MIT switching, cobaltites and manganites have garnered broad interest. Compared to manganites, cobaltites have an additional degree of freedom, the Co ion spin-state transition, and they are 
more likely to exhibit magnetoelectric phase separation \mbox{(MEPS) \cite{khan2012magnetoelectronic,wu2006glassy,senaris1995magnetic}.} MEPS is a spatial coexistence of nano-domains with differing electronic and magnetic phases such as ferromagnetic/metallic domains within an antiferromagnetic/insulating matrix. 
In addition, the magnetic properties of cobaltite thin films are highly sensitive to the film thickness as MEPS can occur more readily at the substrate interface\cite{li2017thickness,torija2011chemically}.
La$_{1-x}$Sr$_{x}$CoO$_{3-\delta}$ (with $\delta$ corresponding to the oxygen deficiency), is a prototypical perovskite cobaltite where the oxygen vacancy and/or Sr-doping concentrations are key control parameters. The P to BM transition (with associated $\sim10^4$ relative change in resistivity) was achieved through changes in temperature \cite{tokura1998thermally}, pressure \cite{lengsdorf2004pressure}, strain \cite{torija2008epitaxial}, Sr \mbox{doping \cite{cheng2015oxygen},} and electrochemical gating \cite{liang2025limits,walter2016electrostatic}.
Nonetheless, the exact mechanisms responsible for ionic-diffusion across the coupled structural and magnetic transition remain an active area of research, and multiple competing pathways are observed \cite{lannerd2026thermodynamics,varela2024exploring}, such as negative hydrogen insertion and oxygen vacancy formation energies.
Recent electrochemical \mbox{experiments \cite{liang2025limits}} have shown fast initial switching on the order of seconds, followed by a slow residual evolution of the resistivity for up to thousands of seconds.
Nano-diffraction imaging has shown that the structural transformation occurs by nucleation and growth of a multitude of domains at the nanoscale \cite{smith2025nanodiffraction,rippy2019}.
In order to experimentally study the slow nano-domain evolution over extended periods of time, we propose to utilize coherent synchrotron X-ray diffraction.

In this work, Bragg XPCS measurements of La$_{0.7}$Sr$_{0.3}$CoO$_3$ (LSCO) thin films explore the spatial and temporal heterogeneity of the first-order P to BM phase transformation.
Specifically, it is found that throughout the transformation, the timescales linked to BM phase formation continuously evolve as a function of time under constant reducing conditions.
We find that the as-grown P phase contains a small amount of preexisting BM domains ($\approx$ 0.2\%), which likely serve as nucleation centers for the first-order phase transformation, in accordance with prior nano-diffraction imaging \mbox{studies \cite{smith2025nanodiffraction,rippy2019}.}
Furthermore, our results show that upon annealing under strongly reducing conditions, the BM domains are governed by two timescales. The faster timescale, on the order of $\sim$1000~s and likely associated with domain growth, is mostly time-independent. The slower timescale, likely associated with domain motion, has a clear time-dependence.
In particular, the slow timescale accelerates by almost an order of magnitude over a 9000~s (2.5~h) time span after the introduction of the thin film to the reducing environment. This result shows that a true equilibrium condition might take much longer to be achieved than could be determined from measurements such as \mbox{X-ray} diffraction (XRD), which are sensitive to the average value for the full volume of the thin film  \cite{rietveld1969profile}. 
Such dynamic heterogeneity, characterized by fast and slow timescales, is suggested to be a universal feature of glass-forming liquids due to the percolation of mobile and immobile domains~\cite{gao2025unified}.

\noindent\textbf{2. Results}\\
A 20~nm thick LSCO thin film grown on a \mbox{(001)-oriented} (LaAlO$_3$)$_{0.3}$(Sr$_2$AlTaO$_6$)$_{0.7}$ (LSAT) substrate was studied as a function of annealing temperature in a P$<$1$\times10^{-3}$~mbar pressure environment. The choice of the LSAT substrates minimizes the epitaxial tensile strain in the as-grown P phase. The measurement temperature was varied from room temperature (25~$^\circ$C) to 550~$^\circ$C in five steps. At each temperature, after an alignment XRD scan, a 2~h long synchrotron XPCS measurement was performed, first on the P(002) peak, then on the BM(006) half-order peak, followed by another XRD measurement.
The X-ray energy was set at 6~keV with a 7$\times$14~$\mu$m$^2$ focused beam and 3$\times$10$^{10}$~ph/s flux.
 \begin{figure*}[ht] 
 \centering   
  \begin{subfigure}[b]{0.48\textwidth}  
  \includegraphics[width=\textwidth]{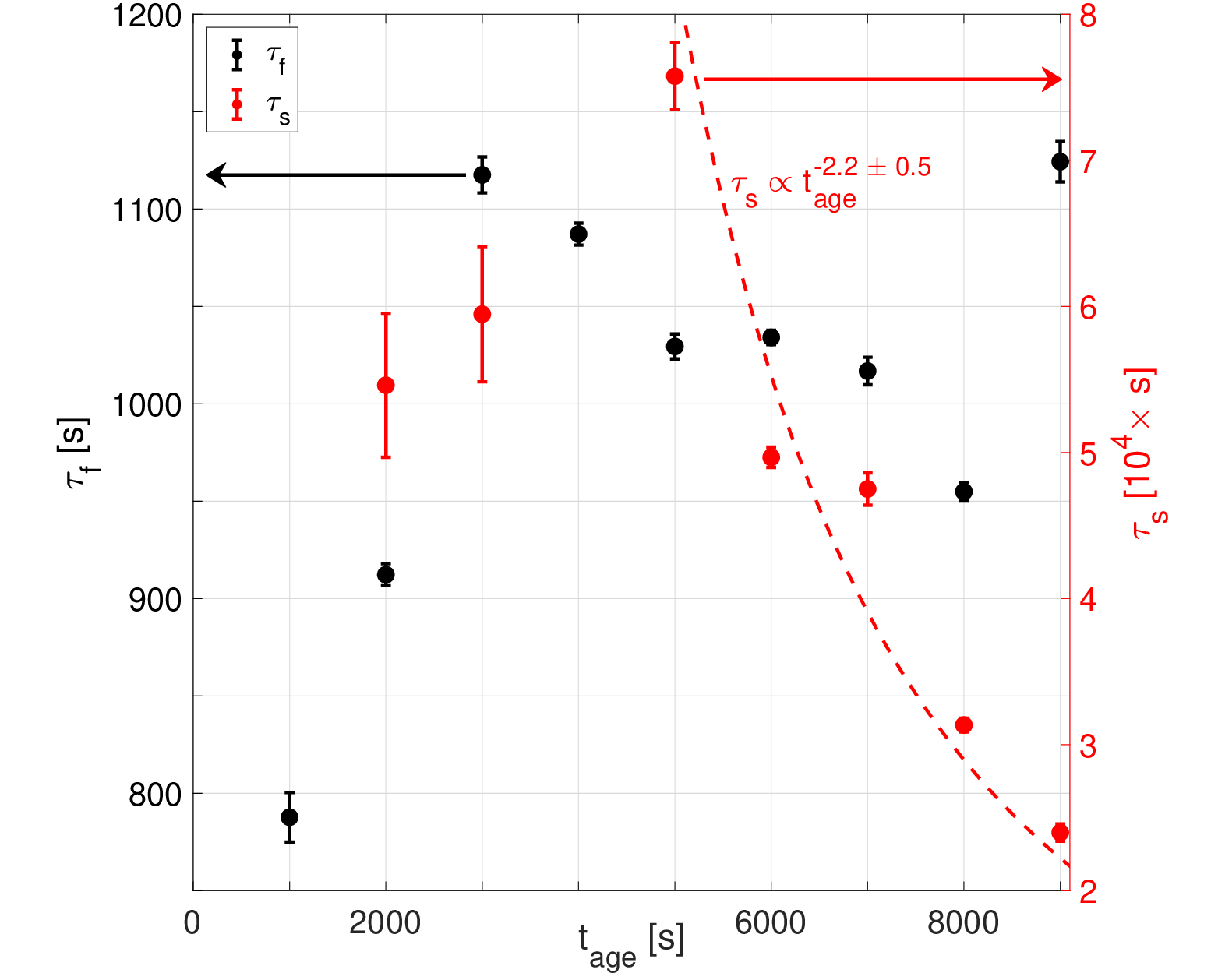}
  \caption{\label{fig_taus}}
  \end{subfigure}
\hfill
  \begin{subfigure}[b]{0.48\textwidth}  
  \includegraphics[width=\textwidth]{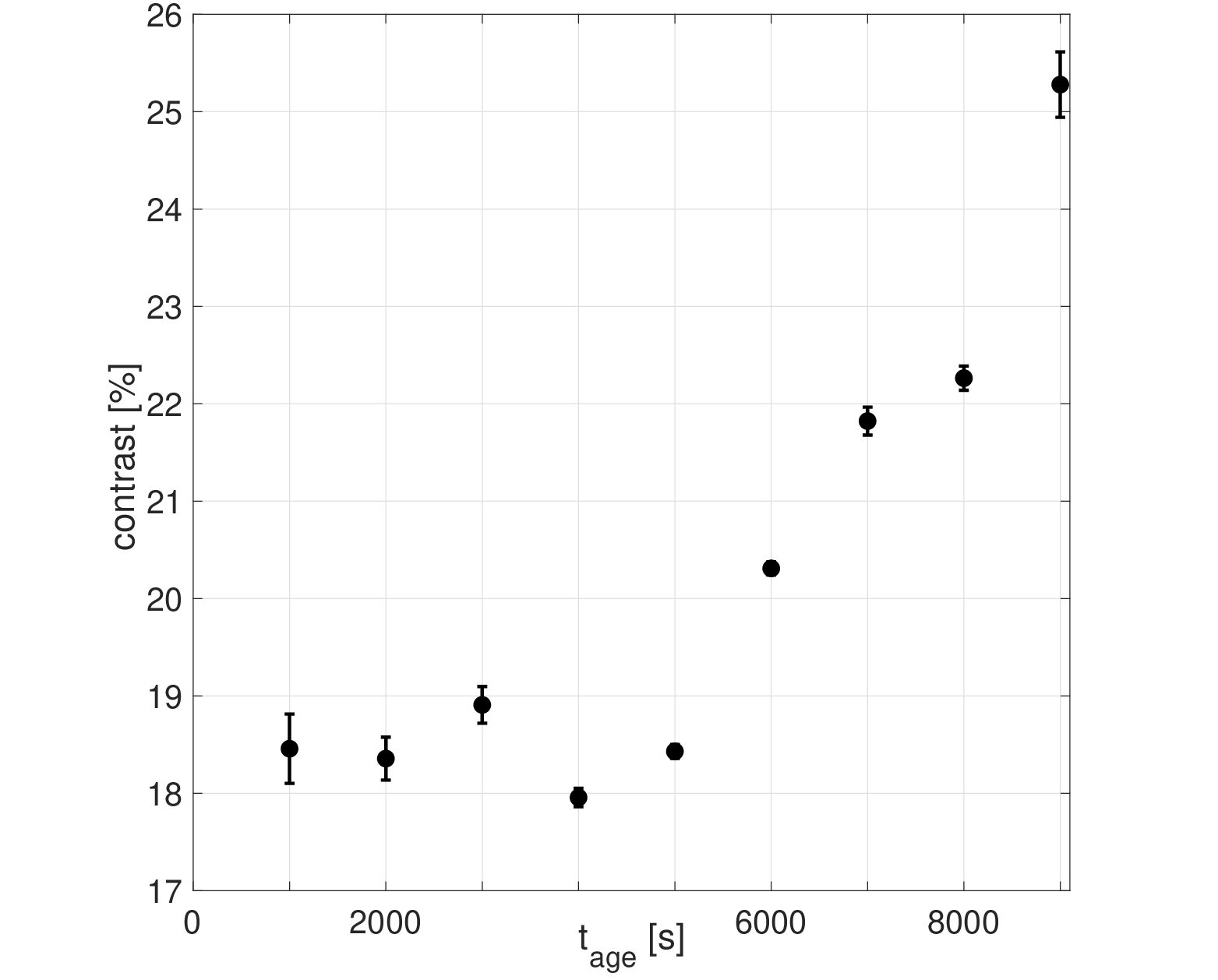}
  \caption{\label{fig_contrast}}
  \end{subfigure}
 \caption{\textbf{Timescales associated with the BM phase at 550~$^\circ$C.} 
 	(a) Slow and fast timescales, $\tau_s$ and $\tau_f$, respectively. $\tau_s$ decreases at t$>$5000~s and can be fit with a power law (dashed line) with exponent -2.2$\pm$0.5.
    \mbox{(b) Contrast} as a function of t$_{age}$ obtained from the second derivative of the fits to the data in Fig.~\ref{fig_g2_fits}. An increase in contrast indicates that a larger percentage of the diffracting phase is dynamic.
    }
\label{figfitresults}            
\end{figure*}

\textit{XRD results:}
As confirmed by synchrotron XRD measurements, shown in Fig.~\ref{fig_setup}c, the LSCO thin film undergoes a first-order phase transformation from the initial P phase to the oxygen deficient BM phase at T$\approx$550~$^\circ$C, as was seen in prior work under similar strongly reducing conditions \cite{chiu2021cation}.
The exact transition temperature depends strongly on the annealing pressure \cite{feng2024hydrogen} and decreases under more reducing conditions. According to our synchrotron XRD measurements, the transition temperature was between 360~$^\circ$C and 550~$^\circ$C.
The transformation is considered to be first-order because the two phases are characterized by a distinct order parameter that is discontinuous at the transition and the XRD measurements did not detect intermediate phases \cite{smith2025nanodiffraction,chaturvedi2021doping}.
The high intensity associated with synchrotron X-rays allows the detection of trace amounts of secondary phases which would not be possible with lab-based XRD systems. While the BM peak is barely visible in the XRD measurements (see Fig.~\ref{fig_setup}c), it is clearly resolved by a 2D detector, with an example shown in Fig.~\ref{fig_setup}a at 240~$^\circ$C. 
The XRD measurements around the BM(006) peak at temperatures below the transition temperature are shown on the right of Fig.~\ref{fig_setup}c.  
At those temperatures, the peak has an approximately constant width and intensity, indicating that these BM domains likely originated during the deposition process.
As the BM structure consists of alternating layers of octahedrally- and tetrahedrally-coordinated B-site cations (see crystal structure diagram in Fig.~\ref{fig_setup}c), the unit cell dimensions in the c-direction are quadrupled relative to the base P phase. As a result, the BM(006) 'half order' peak provide key signatures of the BM phase formation.

The observed speckle patterns in Figs.~\ref{fig_setup}a,~\ref{fig_setup}b, and~\ref{fig_setup}d  are a consequence of the coherent X-rays used for the measurement.
Before the phase transformation, the peak is wide and contains a small number of speckles (Fig.~\ref{fig_setup}a, 240~$^\circ$C), while after the transformation it is narrower and contains too many speckles to be distinguished separately (Fig.~\ref{fig_setup}b, 550~$^\circ$C).
The width of the peak ($\Delta Q_{FWHM}$) is inversely proportional to the average in-plane domain size $d_{avg}$, and the number of speckles is approximately the number of domains in the region illuminated by the \mbox{X-ray} beam \cite{bluschke2022imaging}. 
At 240~$^\circ$C, the mean domain size is \mbox{$d_{avg}\approx$ 60$\pm$5~nm,} and there are on the order of 20 domains in the illuminated region.
At 550~$^\circ$C, the mean domain size is $d_{avg}\approx$ 55$\pm$5~nm.
In the dilute case, below the transition temperature, the smallest \mbox{distance $\Delta Q_{min}$} between the speckles gives the largest distance $d_{max}$ separating domains within the illuminated region, \mbox{here $\approx$ 300~nm \cite{bluschke2022imaging}.}
Figs.~\ref{fig_setup}a and~\ref{fig_setup}b illustrate the real-space domain distributions that can lead to the observed Bragg peaks.
Importantly, below the transition temperature, the ratio of the number of speckles times their mean dimension and the beam spot size gives a measure of the percentage of the sample in the BM phase, approximately 0.2\% at 240~$^\circ$C. 
While no 2D diffraction data were recorded for the BM phase at $T\leq240~^\circ$C, the XRD data shown on the right inset of Fig.~\ref{fig_setup}c show that a small amount of the BM phase existed in the as-grown sample. 

\textit{XPCS results:}
The Bragg XPCS data collected at each temperature were analyzed by computing the usual two-time intensity-intensity correlation function \mbox{$G$(t$_1$, t$_2$) = $\frac{I(t_1) \cdot I(t_2)}{\langle I^2 \rangle_{t}}$,}
where $\langle I^2\rangle_{t}$ denotes the average of the squared intensity over the total acquisition period \cite{shpyrko2014x}.
Thereafter, by taking slices of \mbox{$G$(t$_1$, t$_2$)} with constant time delays t$_2$ - t$_1$ = t, the one-time correlation function $g_2$(t) was obtained.
Finally, the intermediate scattering function $|F(t)|$ was calculated with \mbox{$g_2(t) = 1 + A|F(t)|^2$}, where $A$ describes the beam coherence \cite{sutton1991observation}. 
Fig.~\ref{fig_setup}f shows $|F$(t)$|^2$ obtained on the BM(006) peak (left) and P(002) peak (right) for all temperatures.
The Bragg scattering geometry at large angles ($\theta>$20$^\circ$) probes the in-plane domain characteristics rather than out-of-plane characteristics, because the out-of-plane momentum vector projection $Q_z$ is small and leads to low $Q_z$ sensitivity.
At the transition temperature ($\approx$550~$^\circ$C), the XPCS data reveal dynamics in the BM(006) half-order peak with timescales of the order of 1000~s, consistent with slow ionic diffusion \cite{liang2025limits}. In contrast, for the P(002) peak and at all other temperatures, the dynamics are considerably slower, as visible in Fig.~\ref{fig_setup}f, and might be dominated by beam and measurement setup instability.
Diffraction from a half-order peak is more sensitive to disorder and phase domains, since it originates from reflections on planes separated by twice the lattice constant (smaller Q-vector) and is sensitive to changes on twice the length scale, explaining the difference seen between the P and BM data.
Note that the XPCS measurement on the BM peak is taken at t$>$2~h after setting the temperature to 550~$^\circ$C, such that the transformation is mostly complete (confirmed by stable Bragg peak intensity and width).

The two-time intensity-intensity correlation $G$(t$_1$, t$_2$) obtained from the BM peak at 550~$^\circ$C is shown in Fig.~\ref{fig_twotimes}. 
The diagonal t$_1$ = t$_2$ is trivially $G$(t$_1$, t$_2$) = 1 and is denoted as the aging time t$_{age}$ (black arrow). Taking cuts perpendicular to t$_1$ = t$_2$ (colored arrows) corresponds to calculating the one-time correlation functions $g_2^{t_{age}}$(t) and intermediate scattering function $|F_{t_{age}}$(t)$|^2$ starting at the specific aging time t$_{age}$.
The  $|F_{t_{age}}$(t)$|^2$ corresponding to the colored arrows in Fig.~\ref{fig_twotimes} are shown in Fig.~\ref{fig_g2_fits}, with the dotted lines being fits with a double exponential decay function of the form
\begin{equation}
|F_{t_{age}}(t)| = a*exp(-(t/\tau_f)^\beta)+(1-a)*exp(-(t/\tau_s)^\beta),
\label{exp_fit_eq}
\end{equation}
where $\tau_f$ and $\tau_s$ are the fast and slow characteristic relaxation times, respectively. $\beta$ is the stretching exponent and $a \leq 1$ is a constant.

Noticeable in Fig.~\ref{fig_g2_fits} is that $|F_{t_{age}}$(t)$|^2$ saturates at approximately $\sim$82\% for t$_{age}\leq$300~s, while  for t$_{age}>$300~s a second slow decay becomes visible. The first decay at t$<$1000~s corresponds to the fast timescale $\tau_f$ and the second decay at t$>$1000~s to the slow timescale $\tau_s$ in Eq.~\ref{exp_fit_eq}.
The fitted values for $\tau_f$ and $\tau_s$ are shown in Fig.~\ref{fig_taus} as a function of $t_{age}$, with $\tau_f$ and $\tau_s$ on the left and right y-axis, respectively. 
The fitting parameters $\beta$ and $a$ are found to be t$_{age}$ independent, with \mbox{$\beta$ = 1.6$\pm$0.3} and $a$ = 0.12$\pm$0.02.
A $\beta$ value of 1.5 is indicative of jammed/glassy dynamics \cite{shpyrko2014x}.
The initial fast decay timescale $\tau_f$ has a tendency to increase as a function of t$_{age}$ (corresponding to a slow down in dynamics) and is close to $\tau_f \approx$ 1000~s. The slow timescale $\tau_s$ undergoes a clear decrease (faster dynamics) with t$_{age}$, decreasing by almost an order of magnitude over the 10000~s measurement time, following a power law with \mbox{exponent -2.2$\pm$0.5.}

A measure of the percentage of the BM phase that is dynamic can be quantified by the contrast associated with the fast decay rate $\tau_f$ of the BM(006) Bragg peak. It is obtained by minimizing the second derivative of Eq.~\ref{exp_fit_eq} using the fitting values from Fig.~\ref{fig_g2_fits} and is plotted in  Fig.~\ref{fig_contrast}.
The results show an increase in contrast as a function of t$_{age}$, with a clear increase for t$_{age}>$5000~s. 

\noindent\textbf{3. Discussion}\\
The data presented in Figs.~\ref{figtagedef} and~\ref{figfitresults} might be interpreted as XPCS evidence of nucleation and growth of the BM phase in LSCO under constant reducing conditions (\mbox{T = 550~$^\circ$C} and P$<$1$\times10^{-3}$~mbar).
The XPCS data are taken at t$>$2~h after setting the temperature to 550~$^\circ$C, ensuring that the dynamics are not due to setup thermal instabilities. 
A clear sign that the observed dynamics are linked to the phase transition is that they occur only in the BM peak near the transition temperature (Fig.~\ref{fig_setup}f). 

The age-dependent correlation decay in Fig.~\ref{figtagedef} can qualitatively be explained by a growth and de-pinning process of BM domains. 
Initially, only the motion of domain walls (the transition region between BM and P phases) contributes to the change in the diffraction pattern, leading to the limited 18\% contrast value, meaning that only 18\% of the BM phase within the domains is not static. The non-static part may be due to motion of atoms and defects.
The slow timescale $\tau_s$, which speeds up with t$_{age}$ following a power law with exponent -2.2$\pm$0.5, is linked to de-pinning of BM domains, leading to global domain motion and an increase in the decay contrast visible in Figs.~\ref{fig_g2_fits} and~\ref{fig_contrast}.
The BM domain size is estimated to be 55$\pm$5~nm.
Assuming circular domains, an outer band containing 18\% of the area is 5~nm thick. This 5~nm region corresponds to the domain wall motion (region where the domain grows) and moves at an average speed of $v_d = 6 \pm 0.5 \times10^{-4}$~nm/s, or $v_d = 2\pm0.2$~nm/h (see experimental section for more details).
However, nano-diffraction imaging of the first-order P to BM phase transformation in LSCO/Al and LSCO/Gd heterostructures \cite{smith2025nanodiffraction,rippy2019}, and VO$_2$ \cite{shabalin2020nanoscale} and V$_2$O$_3$ \cite{d2025self,shao2025x} metal-oxides, observed domains to grow by elongation of filaments and to be subject to strain at domain boundaries, which affect the domain shape and ratio of domain wall length to domain size.
Transmission electron microscopy measurements \cite{cui2018direct} have shown large differences between in-plane and out-of-plane domain growth speeds under electrical gating, with lateral speeds up to 30 times faster. Here, because of the Bragg scattering geometry at relatively large angles, only in-plane values are measured and any anisotropy cannot be observed.
The BM(006) Bragg peak before the phase transition, Figs.~\ref{fig_setup}a and~\ref{fig_setup}c (magnification), shows that a small amount of BM (0.2\%) domains exists in the as-grown P phase (see experimental section for more details). 
During the phase transformation, these preexisting domains can act as nucleation centers for the further growth of the BM phase.
These initial domains might be pinned to defects, for example a line defect as observed by nano-diffraction imaging \cite{rippy2019}, which can become mobile upon heating in vacuum when the overall crystal structure changes, resulting in the observed double exponential decay of the correlation functions in Fig.~\ref{fig_g2_fits}. 

The observed timescales, $\tau_f$ and $\tau_s$, are consistent with recent electrochemical switching experiments on LSCO thin film devices \cite{liang2025limits}, where oxygen ion diffusion was observed to set the timescales for the P to BM phase transformation. 
In particular, these experiments found that switching times depend on the square of the LSCO thin film thickness and that there are three timescales linked to the transformation that differ by an order of magnitude from each other. 
They are the time t$_{start}$ for the first BM domains to nucleate, the time t$_{end}$ for the phase-pure BM to be achieved, and the time t$_{off}$ for the source-drain current ON/OFF ratio to reach a value of 5$\times$10$^{4}$.
XPCS measurements can be sensitive to all three timescales and are particularly well-suited to detect the two slower processes, related to t$_{end}$ and t$_{off}$. This is because XPCS measures atomic-scale distribution and dynamics, including motion induced by oxygen diffusion through the phase domains.
The t$_{end}$ and t$_{off}$ times indicate slow BM domain growth (t$_{end}$) and a slower final rearrangement of the domains (t$_{off}$), both expected to be visible with XPCS and indicative of timescales expected from oxygen diffusion in LSCO.
In the electrochemical experiment \cite{liang2025limits}, for a 20~nm thick LSCO thin film, the times were \mbox{t$_{start} \approx$ 50~s,} t$_{end} \approx$ 100~s, and t$_{off} \approx$ 1000~s.
For the thermally driven transformation investigated here, XPCS yields timescales on the order of 1000~s and 10000~s for $\tau_f$ and $\tau_s$, respectively, evolving with t$_{age}$ for up to 10000~s. These values are comparable with the electrochemical switching times, indicating that the same oxygen diffusion-limited dynamics govern both types of switching.
In particular, the slow timescale $\tau_s$ linked to domain motion might be responsible for the difference between t$_{end}$ (phase-pure BM) and t$_{off}$ (ON/OFF ratio 5$\times$10$^{4}$) in devices.

A number of different factors can impact the observed dynamics for the P to BM phase transformation in LSCO thin films. In the current work, LSAT substrates were selected to minimize the epitaxial tensile strain in the as-grown P phase to obtain thin films with high crystalline quality \cite{feng2022strain,chiu2021cation}. However, the lattice expansion that occurs upon transformation to the BM phase will lead to a larger value of epitaxial strain. In contrast, the selection of a substrate with lattice parameters better matched to the BM phase may favor its formation. However, the resulting larger epitaxial strain with the as-grown P phase may introduce more crystalline defects, which can act as nucleation sites for BM domains as well as fast oxygen vacancy diffusion pathways. Because the thin film samples investigated in this work are clamped to their underlying substrates, they are not free to respond to this lattice expansion as in a bulk material. In all these scenarios, the energy landscape for the transformation will be altered and thus affect the domain dynamics. Similarly, the choice of composition \mbox{($x$ in  La$_{1-x}$Sr$_{x}$CoO$_{3-\delta}$)} can influence the A-site disorder, which can impact oxygen vacancy migration barriers, local strain fields, and defect pinning \cite{zhang2022determining}.
For example, disorder could enhance domain pinning and therefore modify the slow de-pinning process identified in this work.
In addition, Sr pileup to the surface and SrO formation has been observed \cite{postiglione2024mechanisms}, which at high coverages may prevent additional oxygen ions from entering the lattice and modify the measured domain dynamics.
The choice of substrate and composition influences the critical conditions for the phase transformation (transition temperature and pressure), and depending on how strongly reducing the conditions are, different transformation pathways, speeds, and microscopic domain structures may be accessed \cite{varela2024exploring,zhang2022determining}.
Thus, the exact domain motion timescales observed by XPCS are expected to depend on sample and experimental parameters. However, the general correlated/glassy dynamic characteristic is linked to oxygen ion diffusion through the lattice and should be less sensitive to the exact material system investigated.

The XPCS results presented here confirm that the first-order P to BM phase transformation in LSCO thin films is diffusion-limited and that while the initial formation of BM domains can be relatively fast ($<$1~s, not measured here), the time to reach a final stable state can be hours, depending on the reducing conditions.
The stretching exponent obtained from fits to Eq.~\ref{exp_fit_eq}, \mbox{$\beta$ = 1.6$\pm$0.3,} confirms the picture of a jammed/glassy system, where the motion of each domain affects all the others.
Such slow time-dependent aging dynamics may limit device speed performance.
It is interesting to note that techniques not sensitive to nanoscale heterogeneity dynamics, such as XRD, are not able to detect slow timescales associated with domain motion once the majority phase is achieved, thus positioning XPCS as an ideal technique for the study of phase transformations.
While here XPCS is used to characterize the annealing of LSCO thin films in vacuum, it would be well suited to complement electrochemical switching experiments of perovskite oxide transistors in order to associate electrical performance with nanoscale domain dynamics. Additionally, by utilizing free electron laser X-ray radiation, the measurements could access timescales as fast as picoseconds \cite{yu2024ultrafast}.

\noindent\textbf{4. Conclusion}\\
Synchrotron XPCS and XRD measurements were used to study a 20~nm thick LSCO thin film under reducing conditions. 
Results show that at T=550~$^\circ$C and \mbox{P$<1\times$10${-3}$~mbar}, the perovskite to brownmillerite phase transformation has dynamical heterogeneity, characterized by two timescales on the order of 1000~s and 10000~s that are time-dependent, \textit{i.e.}, the sample undergoes aging even at times longer than hours, characteristic of glassy materials and possibly posing a limitation to switching speeds.
The two timescales are consistent with oxygen ion diffusion leading to continued brownmillerite phase growth and with de-pinning of brownmillerite domains.
XRD measurements show that a low density ($\approx$ 0.2\% of the sample) of brownmillerite domains are present within the perovskite phase already at room temperature, which may act as initial nucleation centers for the phase transformation.
Altogether, the XPCS results confirm that the phase transformation is oxygen diffusion-limited.
In order to optimize LSCO-based devices, it will be beneficial to measure XPCS on LSCO transistors under operation, to associate domain growth and motion timescales with electrical performance.

\noindent\textbf{5. Experimental Section}\\
\textit{Sample preparation:}
Epitaxial La$_{0.7}$Sr$_{0.3}$CoO$_3$ thin films with 20 nm thickness were grown on (001)-oriented LSAT substrates by pulsed laser deposition using a KrF excimer laser \mbox{($\lambda$ = 248~nm).} The deposition was performed at a substrate temperature of 700~$^\circ$C using a laser fluence of 0.8~J/cm$^2$ and a repetition rate of 1~Hz. The growth was carried out in 0.3~Torr of O$_2$, followed by post-deposition cooling to room temperature in 300~Torr of O$_2$ to ensure proper oxygen stoichiometry.

\textit{X-ray diffraction and photon correlation spectroscopy:}
The \mbox{X-ray} photon correlation spectroscopy experiments were conducted at the 9C beamline of the Pohang Light Source-II (PLS-II) \cite{ham2026upgrade}. 
The beam was focused with Kirkpatrick-Baez mirrors to a spot size of 8$\times$16~$\mu$m$^2$ at 6~keV with a transverse coherence length of approximately 1.6~$\mu$m in both directions with a 3$\times$10$^{10}$~ph/s flux.
Diffraction was measured with an \mbox{EIGER}2 X 1M detector, with 1028$\times$1062 pixels and pixel size 75~$\mu$m, placed at 861.29~mm from the sample at an angle 2$\theta$=22.5$^\circ$ for the BM peak and 2$\theta$=32.8$^\circ$ for the P peak.
With the same setup, the 2$\theta$ angle was swept while recording the total diffraction intensity on the detector for the XRD measurements.

During the measurement on the BM peak at 550~$^\circ$C, the total intensity on the detector was approximately constant with a slight decreasing trend in time, while the full-width at half maximum (FWHM, proportional to the mean domain size) remained constant. 

\textit{Brownmillerite domain percentage calculation:}
For a single domain of (in-plane) radius $R_d$, the measured speckle radius $\Delta Q_s$ is inversely proportional to
\mbox{$R_d = 2\pi/\Delta Q_s$} since the speckle is the Fourier transform of the domain shape. 
In the case of many diffracting domains within the beam (dense case), no single speckle is resolvable (see Fig.~\ref{fig_setup}b) and the FWHM of the Bragg peak is proportional to the average in-plane radius of the domains $R_d = 2\pi/\Delta Q_{FWHM}$. At \mbox{240~$^\circ$C,} it yields \mbox{$R_d \approx$ 60$\pm$5~nm.}
In the case of a small number of domains (dilute case), the number of visible speckles (on the order of 20, see Fig.~\ref{fig_setup}a) is equal to the number of domains within the X-ray beam \cite{bluschke2022imaging}. The FWHM of the underlying Bragg peak gives a measure of the average domain size, and the smallest inter-speckle distance $\Delta Q_{int}$ gives the largest in-plane distance between domains.
Thus, multiplying the number of domains by the average domain area $A_d = \pi R_d^2$ gives the total area $A_{tot} = n\times A_d$ occupied by the phase responsible for the Bragg peak.
Comparing $A_{tot}$ to the X-ray beam size of 8$\times$16~$\mu$m$^2$ gives a BM phase percentage of $\approx$0.2\%.

\textit{Domain growth speed approximation:}
Assuming circular domains, the average domain area is $A_d = \pi R_d^2$ and the moving domain wall area is $A_w = \pi (R_d + R_w)^2 - \pi R_d^2$, where $R_w$ is the distance that the domain wall traveled during the XPCS scan.
The observed correlation contrast is the ratio of the static and dynamic parts of the domain, \textit{i.e.}, 
\mbox{$contrast = A_w/A_d = (R_w + R_d)^2 / R_d^2 -1$,} which decreases with increasing $R_d$ (for a constant $R_w$).
Rearranging the previous expression leads to 
\mbox{$R_w = (\sqrt{contrast+1} - 1) R_d$.}
Since domain growth is attributed to the fast timescale $\tau_f$, and the change in contrast is attributed to domain motion, the initial contrast is taken to calculate the domain growth speed ($\approx$18\%).
The FWHM yields \mbox{$R_d = 2\pi/\Delta Q_{FWHM} \approx$ 55$\pm$5~nm.}
Thus, at 550~$^\circ$C, the BM domains grow by $R_w \approx$ 5~nm over the 9000~s measurement time (Fig.~\ref{fig_contrast}), and at a speed $v_d = 6 \pm 0.5 \times10^{-4}$~nm/s, or \mbox{$v_d = 2\pm0.2$~nm/h.}

\textit{Domain de-pinning rate approximation:}
Domain de-pinning rate is calculated based on the increase in contrast from 18\% to 25\% over a 9000~s period (Fig.~\ref{fig_contrast}), resulting in  8$\times$10$^{-4}$\% of domains de-pinning per second.

\begin{acknowledgments}
We acknowledge the Pohang Accelerator Laboratory (PAL) for provision of synchrotron radiation facilities and we would like to thank Su Yong Lee and Daseul Ham for assistance and support in using beamline 9C. 
This research was supported by the Quantum Materials for Energy Efficient Neuromorphic Computing \mbox{(Q-MEEN-C),} an Energy Frontier Research Center funded by the US Department of Energy (DOE), Office of Science, Basic Energy Sciences, under Award DE-SC0019273.
\end{acknowledgments}

\bibliographystyle{naturemag}
\bibliography{biblio}	
\end{document}